# Contact lens with stretchable distributed-feedback laser for intraocular pressure monitoring

SERGEI A. IVANOV,[1] ILIA M. FRADKIN,[1,*] EKATERINA S. MUSIKHINA,[1] ROMAN V. KIRTAEV,[1] ALEKSANDR A. KHREBTOV,[1] ANDREY A. VYSHNEVYY,[1] ALEXANDER A. MARCHENKO,[1] VALENTIN R. SOLOVEI,[1] ILYA P. RADKO,[1] ALEKSEY V. ARSENIN,[1] AND VALENTYN S. VOLKOV[1]

[1]*Emerging Technologies Research Center, XPANCEO, Dubai Investment Park First, Dubai, UAE*
**fradkinim@xpanceo.com*

**Abstract:** Continuous monitoring of intraocular pressure (IOP) is essential for the diagnosis and management of glaucoma, yet existing clinical methods rely on intermittent in-clinic measurements that miss critical diurnal fluctuations. Smart contact lenses are an attractive platform for continuous IOP tracking, and optical strain sensors are particularly promising thanks to their high sensitivity and natural readout. Most such optical sensors infer strain from the period of a regular structure - a grating or photonic crystal - whose deformation is read out optically. However, the precision of all sensors based on the period of regular structures is fundamentally bounded by the uncertainty principle, which severely limits performance at the millimeter length scales available inside a contact lens. Here, we address this limitation by integrating a distributed-feedback (DFB) laser, based on a surface-modulated ultrathin F8BT dye layer, into a soft polydimethylsiloxane (PDMS) contact lens. Operating at a symmetry-protected bound state in the continuum at the Γ point, the device produces a narrow lasing line whose wavelength shifts directly with grating strain - a quantity not constrained by the spatial uncertainty principle. The DFB structure is fabricated by UV holographic lithography and transferred onto the lens by a simple float-off process, yielding a stretchable, transparent, polymer-compatible sensor that is intrinsically scalable. Tested on a custom artificial eye model whose pressure-induced deformation is comparable in scale to that reported for the human eye, the sensor achieves a sensitivity of 0.027 nm/mmHg and a calibration residual of 1.2 mmHg under phantom conditions, a level relevant to tonometric monitoring, with substantial headroom currently limited mainly by auxiliary readout equipment.

## 1. Introduction

Contact lenses are naturally positioned directly on the ocular surface, which makes them an ideal platform for all kinds of monitoring, treatment, and vision-related interactions. Specifically, contact lenses are routinely used to assist eye tracking[1–11] by increasing precision and simplifying measurements. They are in natural contact with the tear film, enabling monitoring of dry eye syndrome, levels of glucose[12–14], and other biomarkers[15]. Owing to their minimal distance from the iris aperture, contact lenses have unique access to the retina[16,17], hidden deep inside the eye. Such access could be exploited not only for medical purposes but also, for instance, for user identification[18,19]. Moreover, contact lenses are a natural platform for prospective augmented reality devices[20,21], as they allow images to be projected directly into the eye.

Among these many promising applications, one is of particular medical interest: the monitoring of intraocular pressure (IOP)[22]. IOP is a crucial parameter for the diagnosis and management of glaucoma. Currently, however, IOP measurements are performed exclusively by medical personnel, requiring patients to visit a clinic on a regular basis. Such intermittent spot measurements cannot capture the diurnal IOP fluctuations that are critical for accurate diagnosis and effective treatment monitoring. Integrating a sensor into a contact lens could simplify both early diagnosis and continuous monitoring during treatment. This is particularly attractive because contact lenses are already a widely used, easy-to-handle medical device worn by more than 140 million people worldwide[23], with an actively growing user base.

Most currently proposed approaches extract IOP from alterations of the cornea shape. A soft contact lens conforms to the corneal shape; consequently, IOP-induced corneal deformations cause corresponding micro-deformations in the lens that can be detected and measured. The typical corneal deformation corresponding to a pressure change of several mmHg — the clinically required resolution — is estimated to be of the order of 0.1%[24–26]. Because these deformations are not perfectly transferred from the cornea to the lens, the required lens strain measurement precision is even more stringent. Achieving such high precision in a tiny, transparent polymer contact lens is a true challenge. Currently, a number of methods have been proposed and implemented to capture these deformations. SENSIMED, one of the pioneers in this field, has developed a commercially available[27,28] medical device that measures corneal deformations via resistive strain gauges, whose measurements are in turn wirelessly transmitted to the companion device[29–31]. This approach enables continuous IOP monitoring even with closed eyes, but requires an auxiliary component placed on the eyelids and a companion power supply worn on the body. Other implementations of resistive sensing approach[32,33] as well as other promising approaches are mostly at the proof-of-concept stage. Several groups demonstrated radiofrequency-oscillator-based methods[3,14,24,34–37]. Others implement microfluidic strategies that accumulate deformations from relatively large areas and concentrate them in a tiny channel filled with colored liquid[38–40]. Nevertheless, one of the most promising directions is the optical measurement approach, since it typically offers extremely high sensitivity in a small footprint and provides clear and natural readout procedures. In particular, deformations have already been measured via color changes in a photonic crystal[25]. Others utilize the moiré effect between two gratings deformed in slightly different ways, making tiny deformations visible at a macroscopic level and detectable with an ordinary camera[41–43]. There is also the option to directly measure the change in diffraction angle to estimate deformations[44]. These approaches differ widely in what they measure and in how far they have been validated, and the only commercially available one does not return an absolute pressure at all. A comparison of contact-lens pressure sensors, giving for each the sensing principle, the reported response, the level at which it has been validated and its principal limitation, is collected in Supplementary Note 5.

Mentioned optical approaches are directly or indirectly based on a change in the period of regular structures. Their precision is therefore fundamentally limited by the uncertainty principle, which we analyze quantitatively in the Results section. There is a strong demand for approaches that can surpass this limit.

In addition to these fundamental limits, practical measurement conditions impose further constraints. The achievable precision is typically governed by the relative positioning of the measurement components, and theoretical bounds can only be approached under consistent, stable laboratory conditions. In particular, grating-based measurements are highly sensitive to the relative position of the light source and photodetector. It is thus important to develop a measurement scheme whose precision does not depend on the alignment of the optical components.

Beyond raw sensitivity, any contact-lens-integrated optical sensor must also satisfy a stringent set of fabrication requirements that are rarely discussed but eventually dictate which approach is practically feasible and can ever go beyond the prototype stage. An integrated sensor must, (i) be compatible with soft, transparent, stretchable polymer substrates such as hydrogel or polydimethylsiloxane (PDMS), (ii) tolerate the curved geometry of the lens, (iii) feature sub-wavelength periodicity over a millimeter-scale length to maximize the number of working periods within the limited lens area, and (iv) be manufacturable by a scalable, low-cost process, since contact lenses are inherently disposable. These requirements are difficult to reconcile: established sub-wavelength patterning routes such as electron-beam lithography or focused-ion-beam milling are expensive and not suitable for fabrication of centimeter-scale structures. Scalable sub-wavelength patterning is nevertheless available through well-established routes such as nanoimprint lithography, which is naturally suited to the mass production of optical structures on polymer substrates; UV holographic lithography, used here, complements it as a fast, mask-free, large-area route that is particularly convenient for prototyping. The remaining challenge is therefore not the patterning step itself, but combining sub-wavelength period over a millimeter-scale length, compatibility with a soft and curved substrate, and integration with a soft polymer contact lens within a single process flow. A fabrication process that simultaneously delivers sub-wavelength period over millimeter-scale length, polymer compatibility, and scalability is therefore as much a part of the requirement as the device sensitivity.

In this work, we address these challenges by implementing a distributed-feedback (DFB) laser based on a surface-modulated ultrathin dye layer consisting of the fluorene copolymer poly(9,9-dioctylfluorene-alt-benzothiadiazole) (F8BT), which serves as the core of the waveguide for the lasing mode. This compact, stretchable structure is fabricated by UV holographic lithography of a positive photoresist followed by replication into the F8BT waveguide and a float-off transfer onto PDMS - a route that simultaneously delivers sub-wavelength period, millimeter-scale aperture, and native compatibility with soft and stretchable polymer substrates, without requiring electron-beam lithography or focused-ion-beam milling. While holographic lithography is well suited to rapid prototyping, the sensor architecture itself is fully compatible with mass production, as its sub-wavelength grating can be replicated at scale by well-established nanoimprint lithography. The structure is therefore directly suited for integration inside a contact lens. The emission wavelength of such a structure is entirely determined by the period of the underlying grating, thus providing direct sensitivity to mechanical deformations. Most importantly, however, its laser emission linewidth is disproportionately small relative to the size of the grating, which enables an extremely precise measurement of the emission wavelength and, consequently, of the deformations and IOP value. Based on an in-house developed artificial eye model and a contact lens, we demonstrate a sensitivity of 0.027 nm/mmHg and a calibration residual of 1.2 mmHg for a sensor placed on the surface of the lens. This level of performance is of the order required for tonometric monitoring, although it is demonstrated here under phantom conditions only. Furthermore, the proposed device has a strong potential for further improvement, paving the way for continued development and full integration of the sensor inside contact lenses.

## 2. Methods

### *2.1 Design of artificial sclera–cornea membrane and contact lens*

To evaluate the pressure-sensing performance of the IOP sensor, we required an artificial eye model whose pressure-induced shape changes at least qualitatively resemble those of a real eye. Although the human eye is a mechanically complex structure comprising multiple tissues with distinct properties, a simplified model is sufficient to mimic eye properties relevant for contact lenses. The primary mechanical characteristic of interest is corneal deformation. To reproduce the sensitivity and overall shape change of the cornea under pressure, we designed a geometry that qualitatively mimics the main anatomical features of the eye. The model contains a central corneal region with a radius of 7.8 mm and a lateral scleral region with a radius of 11.0 mm (see Fig. 1 (a)). The membrane thickness was 450 µm in the corneal region and 900 µm in the scleral region. The limbal transition between these zones was smoothed with fillets, and a dedicated skirt at the scleral edge was included for convenient mounting.

The contact lens design was defined to be complementary to the eye model. Its inner surface precisely matches the outer surface of the artificial eye, while its outer surface is spherical with a 9.0 mm radius. The lens thickness at the central point was set to 400 µm.

Mechanical deformations of both the eye model and the contact lens were simulated using the Structural Mechanics Module of COMSOL Multiphysics. Pressure was applied gradually to the inner surface of the membrane. At each step, the solution from the previous step was used as the initial guess for the next one, which accelerated convergence for the nonlinear deformations of the soft polymer structure. Both parts were assigned the PDMS properties: Young's modulus E = 1.47 MPa and Poisson's ratio ν = 0.45, which allowed us to reproduce the experimentally observed deformations. As discussed in detail in Supplementary Note 1, these material properties enable the model to accurately reproduce the deformation behavior of the experimental sample; furthermore, the stiffness of the artificial eye corresponds well to that of a real eye. We emphasize, however, that this remains a strongly simplified model: it is designed to reproduce the scale of the pressure-induced deformations and the overall change of shape, and not the many other aspects of a real eye. Validation under more realistic conditions - ex vivo animal eyes, in vivo animal studies and, ultimately, human trials - will therefore be a necessary step of the further development.

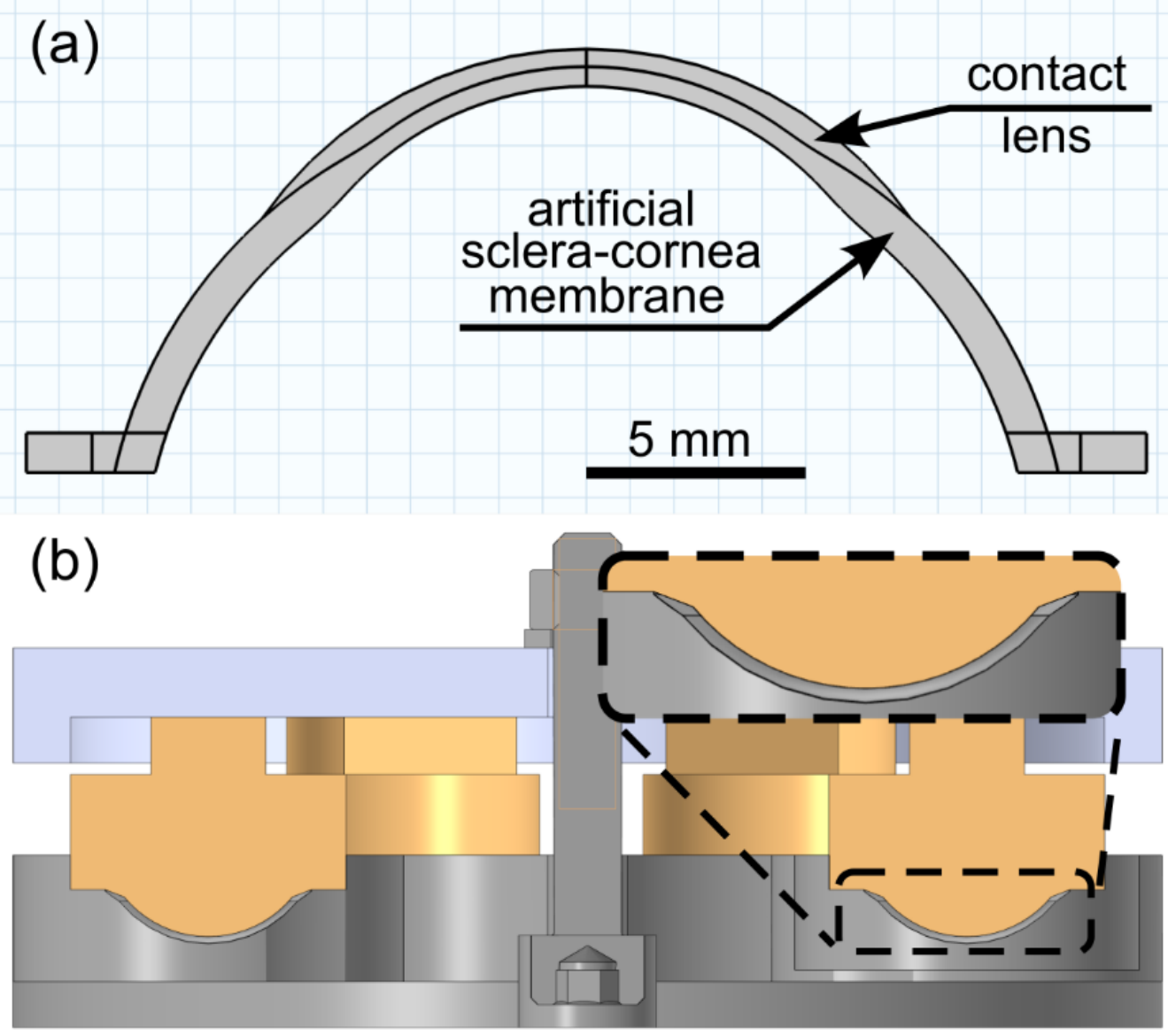

Fig. 1. (a) Cross section of artificial sclera-cornea membrane and contact lens. (b) Scheme of the molds applied for lens fabrication. Similar ones are used for membrane fabrication.

### *2.2 Numerical modeling of optical band structure*

To simulate the optical properties of the DFB laser structure, we used the Fourier Modal Method (FMM)[45] in the scattering-matrix formulation, also known as Rigorous Coupled-Wave Analysis (RCWA)[46]. This is a specialized numerical technique for efficient treatment of layered periodic structures. We calculated reflection spectral maps to design the structures to reveal the dispersion of quasi-guided modes and their hybridization.

### *2.3 Fabrication of the PDMS artificial sclera–cornea membrane and contact lens*

Both the eye-model membrane and the soft contact lens were fabricated from two-component PDMS (Sylgard 184, 10:1 base-to-curing-agent ratio by weight). Eye membranes were obtained by thermal casting of the degassed PDMS in custom 3D-modelled molds (see Fig. 1 (b)), whose detailed mechanical design ensured dimensional reproducibility and interchangeability between samples. Contact lenses were cast in modular eight-cavity titanium mold set producing lenses of 16 mm diameter by curing at 70 °C for 2.5 h.

### *2.4 Fabrication of DFB laser*

Our fabrication route was designed to satisfy, in a single process flow, the four requirements identified in the Introduction: sub-wavelength periodicity, millimeter-scale aperture, compatibility with soft polymer substrates, and scalability. UV holographic lithography produces the master grating in a single interference exposure over the full aperture, intrinsically free of stitching errors and eliminating the throughput bottleneck of serial techniques such as electron-beam lithography or maskless laser writing; it routinely covers apertures of several millimeters to centimeters with sub-wavelength period - the exact regime relevant to a contact-lens sensor. The use of a positive-tone photoresist provides direct control of the duty cycle through exposure dose and development time, which is essential for tuning the coupling strength of the DFB structure. Replication into the F8BT waveguide via a sacrificial PEDOT:PSS layer enables groove-depth control through a simple spin-coating step. Subsequent float-off transfer is maskless, low-temperature, and substrate-agnostic, allowing the complete laser film to be deposited onto a soft, curved PDMS lens in a single step. It can be performed without the additional imprint-and-transfer stages that would otherwise be required to pattern the active layer on the final substrate. As a result, the entire active element of the sensor can, in principle, be fabricated at wafer scale and transferred onto contact lenses individually - a workflow that is naturally compatible with the disposable nature of contact-lens products.

For this work, gratings with a 5 mm aperture were produced. We used mr-P 1201LIL positive photoresist (micro resist technology GmbH) together with an AZ Barli II 200 bottom anti reflective coating (BARC, MicroChemicals GmbH). The BARC thickness was optimized for the laser wavelength, and the photoresist was spin coated on top to a thickness of 100 nm. Spectral sensitivity of the given photoresist is in the range of 330-450nm. For interference patterning, we employed a single mode He–Cd laser (Kimmon IK3501R G, $TEM_{00}$, 50 mW, 325 nm). The resulting grating had a groove depth of 100 nm, equal to the resist layer thickness. Because the DFB grating is replicated from a single master, the grating period is common to all devices and does not contribute to the device-to-device wavelength spread. The only fabrication-induced variation is the F8BT film thickness, which shifts the wavelength through the effective index. For a calibrated spin-coating curve (film-thickness spread ≈ ±5 nm) this gives a device-to-device spread of $\Delta\lambda \approx 1.7$ nm.

The DFB structure was formed by replicating the master grating into the F8BT waveguide layer. First, the master was treated with a soft air plasma to make the surface hydrophilic, then spin-coated with a ~20 nm sacrificial layer of PEDOT:PSS (Clevios P VP AI 4083). The thickness of this layer directly determines the groove depth of the replicated grating: we observed that the groove depth is reduced during transfer by the thickness of the sacrificial layer. Next, a solution of F8BT (Lumtec, Mw = 10 000 - 20 000) in toluene (25 mg/ml) was spin-coated onto the PEDOT:PSS layer; the dye thickness was 306 nm (see Supplementary Note 3 for the dye spin curve). At this stage, the DFB structure is complete and can be transferred to the desired substrate by a float-off technique[47]. This step is critical for contact-lens integration: it is performed at room temperature, requires neither pressure nor adhesion promoters, and is agnostic to the curvature, softness, and chemistry of the receiving substrate, which makes it directly compatible with soft PDMS lenses and, by extension, with potential roll-to-roll or wafer-scale production of the active element prior to lens assembly. Characterization of the grating was performed by atomic force microscopy (AFM), utilizing Oxford Instruments Asylum Research Inc. Cypher S Standard AFM microscope equipped with a Oxford Instruments TRACK 300 probe (300 kHz, 37 N/m).

### 2.5 Artificial eye model

A compact, battery-powered artificial eye model was developed to enable IOP sensor testing (see Fig. 2). Pressure is generated by a medical syringe driven by a stepper motor with a screw shaft (90 mm, 0.8 mm pitch). Pressure is independently monitored with an external manometer (Omega Engineering PX3005-005AI).

The eye module itself consists of a sealed base containing a gas-filled pressure chamber, a PDMS artificial sclera clamped by a contoured plate with an O-ring seal. A PDMS contact lens carrying the DFB IOP sensor is coupled to the sclera via silicone oil that mimics the tear film.

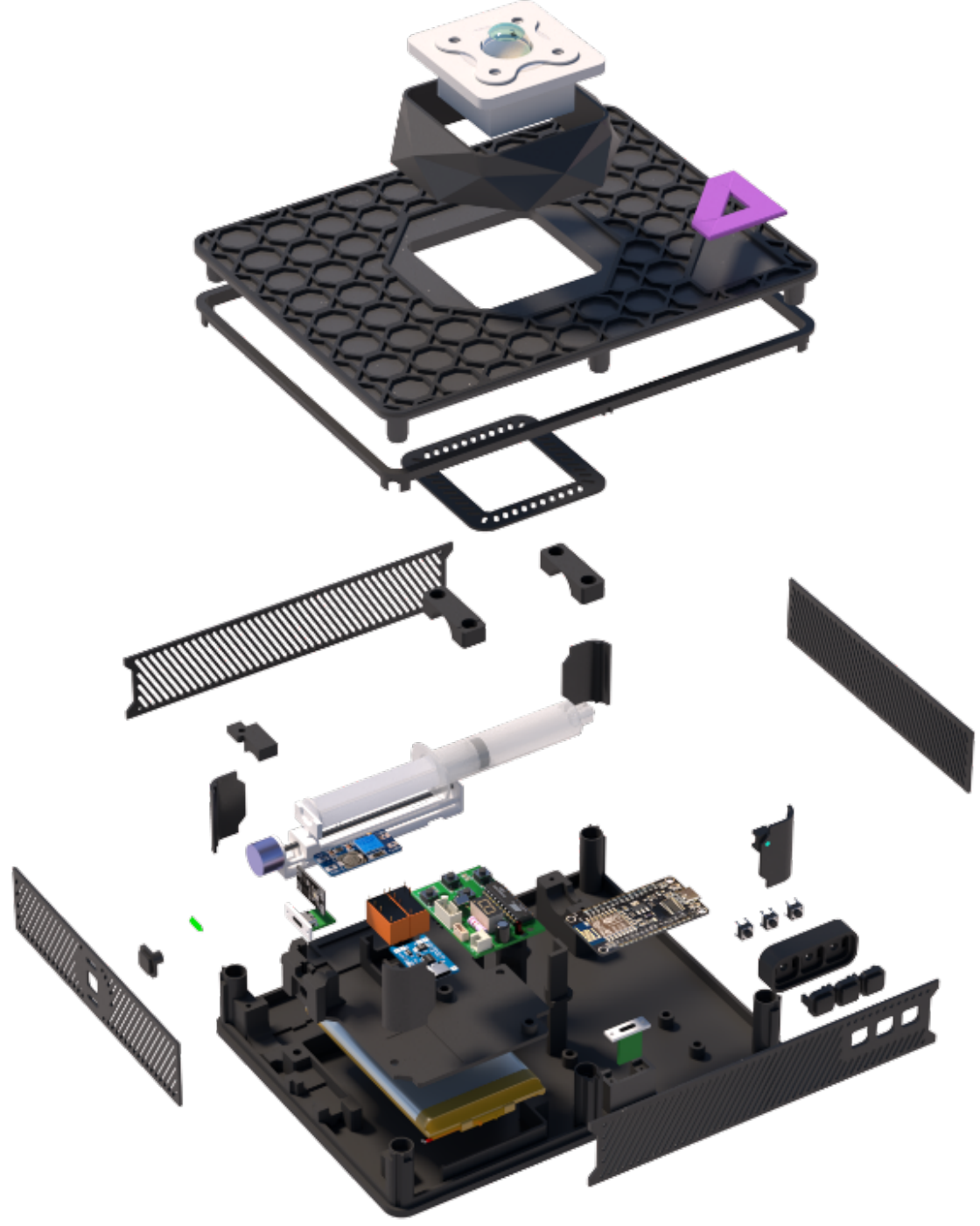

Fig. 2. Exploded view of the assembly of the Eye Model prototype for the IOP sensor.

### 2.6 Measurements

The laser film was placed on top of a PDMS lens. The DFB laser was positioned on the limbal zone of the lens, where the deformations are strongest, with the grating lines aligned radially so that the sensor responds to tangential strain. The contact lens with the transferred DFB laser was then placed onto the eye model.

The DFB structure was optically pumped by a tunable DPSS laser equipped with an optical parametric oscillator (Q-TUNE-F10-APU2, Quantum Light Instruments). A pump wavelength of 510 nm was chosen to reduce thermal effects and optimize energy transfer to the dye. The pump beam was first focused on a diffuser to produce a uniform intensity distribution, then collimated with another lens and focused with a microscope objective (Olympus Plan Achromat). Because of the high pulse energy available, the pump was focused to a point ≈ 17 mm in front of the sample, so that at the sample plane the beam had diverged to a spot of approximately 5 mm. Lasing was observed at all pump energies measured, down to 87 µJ per 8 ns pulse (see Supplementary Fig. S5), so the lasing threshold of the present devices lies below this value, corresponding to a fluence of ≈0.44 mJ/cm². The measurements reported here were performed above threshold, at a pump fluence of ≈1 mJ/cm² (≈200 µJ per pulse). The large spot was chosen deliberately to keep the pump fluence low and does not reflect a need to illuminate the full sensor area, since threshold is governed by fluence rather than total energy. The laser emission was separated from the pump beam by a long-pass filter (Thorlabs DMLP505), and the emitted laser wavelength was measured with a fiber-coupled spectrometer (HR-4UVV250-10, Ocean Optics), while the pressure applied to the eye model was monitored simultaneously with the manometer. The eye model was pressurized up to 45 mmHg above ambient pressure. A single loading–unloading cycle was performed on one device to verify the stable performance of the DFB sensor and the reversibility of dye-film stretching.

The calibration comprised ten points in total — five acquired during pressure application and five during release. The pump repetition rate was 10 Hz, and the spectrometer integration time was 100 ms, so that each recorded spectrum contained a single pump pulse and corresponds to a single generation event. Each plotted point is therefore a single-pulse spectrum recorded simultaneously with the manometer reading. At every pressure level, 20–30 pulses were monitored to confirm spectral stability, but the points were not time-averaged: the calibration residual is dominated by the pressure side, i.e. by the manometer accuracy and the hysteresis between the branches (Section 3.4), rather than by the pulse-to-pulse scatter of the fitted wavelength, so averaging would not reduce it, while reading the wavelength and pressure at the same instant avoids broadening introduced by the slow drift of the pressure baseline described below.

The pressurized chamber is not thermally isolated: although the gas is enclosed beneath the artificial membrane, it exchanges heat with the surrounding air through the membrane and the chamber walls. Compression of the gas raises its temperature slightly, and this heat dissipates to the surroundings during the dwell at each pressure level. We therefore expect the gas on the release branch to be cooler than on the application branch, so that at nominally equal set-points the gauge pressure is lower, and near the end of the release branch it falls slightly below ambient. We attribute the negative gauge values in the calibration (Section 3.4) to this thermal drift of the pressure baseline rather than to an artifact of the wavelength measurement: the wavelength of the last calibration point is correspondingly lower than at the start of the cycle, so the sensor follows the sub-ambient pressure. We suppose that the same drift also contributes to the hysteresis between the branches; other possible contributions are discussed in Section 3.4. The plotted values are raw manometer readings; no offset correction was applied and no suction was used.

## 3. Results

### 3.1 Precision limit of period-based sensors

Let us consider the general case of a sensor based on a periodic structure. As already mentioned, the precision of optical measurement approaches that rely on detecting a change in the period of regular structures is fundamentally limited by the uncertainty principle $\Delta k \cdot \Delta x \approx 2\pi$. When applied to light diffraction on a grating, this expression is equivalent to the Rayleigh criterion. Considering deformations of a grating in real space, the principle can be interpreted as follows: the smallest measurable change corresponds to a change, by one unit, in the number of periods fitting within the grating length. In this way, the minimal detectable strain set by the uncertainty principle (UNP) can be estimated as

$$\varepsilon_{min}^{UNP} \approx \frac{a}{L} = \frac{a}{Na} = \frac{1}{N},$$

where $a$ is the grating period, $L$ the total length of the grating, and $N$ the number of periods. This expression shows that the only parameter that matters is the number of periods. In the context of a contact lens, the total grating width $L$ is limited to a few millimeters. Moreover, the non-homogeneity of the deformations can significantly reduce the effective "correlation length" of the grating. Therefore, the number of working periods can be increased only by reducing the period, which ultimately leads to sub-wavelength photonic-crystal structures or diffraction gratings. One can estimate that a grating with $N \approx 10^2$–$10^4$ periods can realistically be integrated inside a lens, corresponding to a deformation precision limit of $\varepsilon_{min}^{UNP} \approx 10^{-4}$–$10^{-2}$. Given that the reported relative change of corneal radius with pressure, $\frac{1}{R}\frac{dR}{dP}$, is approximately (1–4)×$10^{-4}$ $mmHg^{-1}$ [24–26], the upper limit of $10^4$ periods may suffice to resolve IOP variations of several mmHg — the generally required precision. In practice, however, various factors and experimental errors not accounted for in such crude theoretical estimates prevent one from reaching the theoretical limit. It is therefore necessary to surpass this uncertainty limit and further reduce the detection threshold.

One way to approach the practical limit is to exploit prior knowledge about the system that is not captured by the bare uncertainty relation. For instance, knowledge of the angular or spectral profile of a light beam can allow its center to be located more precisely than the nominal uncertainty limit. Similarly, knowing that the shape of a moiré pattern is sinusoidal can greatly improve the accuracy of period estimation[1]. However, this approach is also limited by the noise level of the structure, depends on the post-processing algorithms employed, and can be reliably estimated only in practice. Therefore, a physics-based measurement method that can surpass the uncertainty limit remains needed.

Another way to achieve this is to measure a different quantity that is not restricted by the uncertainty principle applied to the momentum–coordinate pair. In particular, using a DFB laser as the sensor element allows us to measure the emission wavelength instead of the period or angle. The main advantage is that the emission wavelength is disproportionately narrowed, whereas the sensitivity of the wavelength shift to deformations remains generally unchanged. Moreover, this scheme also benefits from strong prior knowledge: the source emits light at a single frequency with a known line shape, so its center can be determined very precisely. Let us examine this effect using the structure employed in our study.

### *3.2 Resonator design and band structure*

The fabricated slab waveguide consists of a modulated dye layer on the PDMS surface, as shown in Fig. 3(a). From atomic force microscopy (AFM), the grating period is 335 nm, and the profile is nearly harmonic (sinusoidal) with a modulation depth of 75 nm. The total thickness of the structure, including the modulated region, is about 306 nm. In our numerical simulations, we used $n_{PDMS}$ = 1.43 for PDMS, and for the dye layer we used literature data[48], which includes substantial dispersion in the spectral range of interest. At 570 nm, the dye refractive index is $n_{dye}^{570} \approx 1.84$, so it effectively acts as the waveguide core.

Such a structure supports both TE and TM waveguide modes. Due to the symmetry of the structure, guided modes of different polarizations do not couple with each other, but modes of the same polarization interact through the grating and form hybrid modes in the corresponding anti-crossing regions. We focus on the hybridization of TE modes because the grating provides stronger coupling for this polarization, which is highly advantageous for lasing. The band structure for this polarization is shown in Fig. 3(b) as the squared absolute value of the reflection coefficient $\left|r_{ss}\right|^2$. Above the light cone in air, this quantity coincides with the ordinary intensity reflection coefficient; however, unlike the latter, it is also defined below the light cone, allowing the dispersion of the eigenmodes in that region to be resolved.

The diagram clearly shows the fundamental $TE_0$ mode just below the PDMS light cone. Surface modulation of the dye layer folds these modes into the first Brillouin zone, where they undergo anti-crossing at its boundary (the anti-crossing points are marked by the corner brackets). Since no diffraction channels are open in this region, both hybrid modes exhibit theoretically infinite quality factors and may compete uncontrollably for lasing. For this reason, we focus instead on the hybrid modes above the light cone: typically, one of them is bright with a relatively broad linewidth, while the other is dark with a narrow line. In particular, multiple anti-crossing points are observed between the counter-propagating $TE_0$ and $TE_1$ modes. Our primary interest, however, lies in the hybridization of the fundamental $TE_0$ modes, whose field is mainly localized within the gain medium of the dye layer. Moreover, the period and thickness of the structure were chosen so as to position the anti-crossing wavelength at ~570 nm (Fig. 3(c)), close to the peak of the dye spontaneous-emission spectrum (see Supplementary Note 2). Because hybridization occurs between identical modes, they intersect exactly at the Γ point (yellow corner brackets). Furthermore, due to symmetry and perfect destructive interference of the out-coupled fields in the far-field, the dark hybrid mode of the ideal infinite structure becomes infinitely narrow - a so-called symmetry-protected bound state in the continuum (BIC) [49,50]. In the finite, absorbing device studied here, this state becomes a high-Q quasi-BIC that provides distributed feedback for surface-normal lasing. The high-Q dark mode naturally wins the mode competition with the bright one, ensuring stable lasing and simplifying practical measurements. In addition, operation at the Γ point corresponds to emission close to the normal to the grating, which facilitates light collection and detection. Strictly speaking, the mode of an ideal infinite lattice is decoupled from the far field at Γ; in a finite resonator the eigenmodes are quantized and mixed at the edges and at imperfections, and are therefore spread over a range of in-plane wavevectors $\Delta k \approx 2\pi/L_{eff}$, where $L_{eff}$ is the effective length over which the grating remains coherent. This restores a weak coupling to the far field, while the associated divergence in the plane of the grating vector, of the order of $\lambda/L_{eff}$, stays small; along the grating lines, where the one-dimensional grating provides no feedback, the emission diverges strongly, so that the far field has the form of a narrow stripe.

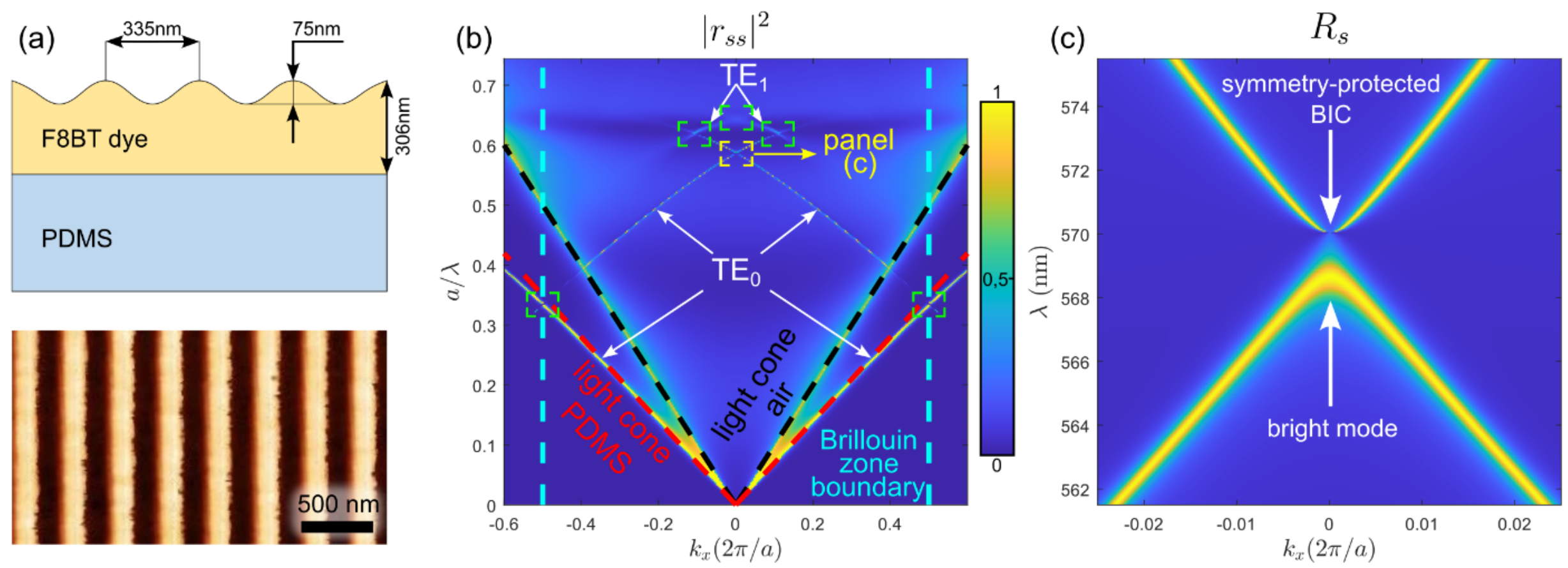

Fig. 3 (a) Schematic of the DFB laser based on a slab waveguide: a dye layer with a harmonically modulated surface on a PDMS substrate, and an atomic force microscopy image showing the surface modulation. (b) Band structure of the grating revealing multiple TE-modes hybridizations. (c) Enlarged reflection spectrum near the Γ point showing hybridization of the $TE_0$ modes, leading to a high-Q dark mode that provides feedback for DFB laser emission.

### *3.3 Sensitivity and linewidth*

We now discuss the two most important characteristics of our structure: the sensitivity of the lasing wavelength to deformation and the mode linewidth. As seen in Fig. 3(c), the mode interaction results in a hybrid-mode splitting of only ~2 nm. The DFB laser wavelength $\lambda_{DFB}$ can therefore be well approximated by the intersection wavelength of the uncoupled modes. The sensitivity of this wavelength shift is essentially the same as that of conventional periodic or diffractive structures, since it is governed primarily by the change in period. The Bragg condition requires the mode wavenumber to satisfy $k_{TE}(\lambda_{DFB})=\frac{2\pi}{a}$. When the grating is stretched, its period increases, leading to a corresponding increase in the lasing wavelength. At the same time, stretching in one direction induces Poisson compression of the layer thickness, which tends to decrease the wavelength and partially suppresses the primary effect. However, owing to the large uncertainty in the actual deformation field and in the mechanical properties of the dye layer, we neglect this secondary effect in our qualitative estimation.

The sensitivity can be estimated by applying the chain rule, assuming the uncoupled mode dispersion does not depend on the deformation:

$$\delta\lambda_{DFB}=\frac{\partial\lambda_{DFB}}{\partial\omega_{DFB}}\frac{\partial\omega_{TE}}{\partial k_{TE}}\frac{\partial k_{TE}}{\partial a}\delta a.$$

Evaluating each derivative yields

$$\delta\lambda_{DFB}=-\frac{\lambda_{DFB}^2}{2\pi c}\cdot v_{TE}^{gr}\cdot\left(-\frac{2\pi}{a^2}\right)\delta a=\frac{\lambda_{DFB}^2}{a}\frac{1}{\beta_{TE}^{gr}}\varepsilon_{lens},$$

where $v_{TE}^{gr}$ is the group velocity, $\beta_{TE}^{gr}=\frac{c}{v_{TE}^{gr}}$ is the group index of the $TE_0$ mode, and $\varepsilon_{lens}=\delta a/a$ is the local strain of the lens. Finally, the wavelength response to pressure is

$$\frac{\partial\lambda_{DFB}}{\partial P}=\frac{\lambda_{DFB}^2}{a}\frac{1}{\beta_{TE}^{gr}}\frac{\partial\varepsilon_{lens}}{\partial P}.$$

Inserting the numerical values gives $\frac{\partial\lambda_{DFB}}{\partial P}\approx\frac{570^2 nm^2}{335 nm}\cdot\frac{1}{2.8}\cdot(1-4)10^{-4}\frac{1}{mmHg}\approx 0.03-0.14\frac{nm}{mmHg}$. The group index extracted from Fig. 3(c) significantly exceeds the phase index, primarily due to the strong dispersion of the dye in this spectral range. The resulting value is an overestimate, since Poisson compensation has been neglected; nevertheless, it provides a reasonable order-of-magnitude estimate. We note that the strain per unit pressure used here is a literature value for the human cornea. For the artificial eye the corresponding quantity follows directly from the finite-element model of Section 2.1: at the position of the sensor on the lens it is (1.5–2.9)×$10^{-4}$ $mmHg^{-1}$, depending on whether the lens slides on the membrane or adheres to it (Supplementary Note 1), which lies within the literature range. With $d\lambda/d\varepsilon \approx 346$ nm this corresponds to 0.05–0.10 nm/mmHg, two to four times the measured sensitivity (Section 3.4), which is the direction expected from the neglected Poisson compression and the imperfect transfer of strain into the dye film.

Sensitivity alone is insufficient; the pressure measurement error $\sigma_P = \sigma_\lambda / \left( \frac{\partial \lambda_{DFB}}{\partial P} \right)$ also depends on the uncertainty in the wavelength determination, $\sigma_\lambda$. In the ideal case, the wavelength measurement error is some fraction of the laser linewidth, which is itself much narrower than the passive-resonator linewidth but still linked to it. The resonator linewidth is limited by at least two main contributions: (i) edge losses arising from the leakage of guided modes at the facets of the finite-length grating, which decrease as the grating size increases, and (ii) intrinsic losses due to out-of-plane radiation and material absorption. The latter contribution can be assessed from the linewidth of the modes in the underlying infinite structure shown in Fig. 3(c). Outside the hybridization region, the linewidth is finite, but it narrows rapidly as the wavevector approaches zero and the coupling between modes becomes significant. We cannot, however, operate exactly at the Γ point, owing to a fundamental limitation imposed by the uncertainty principle: a finite-sized resonator cannot support modes with strictly zero wavevector ($k = 0$). Nevertheless, the minimum wavevector set by the uncertainty principle for a millimeter-scale laser is still very small, $\Delta k = \frac{2\pi}{L} = \frac{a}{L}\frac{2\pi}{a} \approx \frac{335\,nm}{1\,mm}\frac{2\pi}{a} \approx 3 \cdot 10^{-4}\frac{2\pi}{a}$ and lies well within the anti-crossing region of the strongly coupled modes, where high-quality modes are accessible. In other words, the typical grating length is sufficient to fully exploit the feedback from mode coupling, which is the essential principle of any DFB laser.

We additionally note that such a narrow laser line is possible because the uncertainty principle now governs the time–frequency pair rather than the position–momentum pair. The hybrid mode lifetime is greatly enhanced due to reduced far-field radiation losses, and its group velocity is strongly decreased by mode hybridization, allowing the mode to circulate longer before losing energy at the edges. Consequently, the spectral linewidth becomes correspondingly narrow, in full agreement with the uncertainty principle.

### *3.4 Pressure measurements on the artificial eye*

In our experiments, the DFB laser was placed on the lateral part of the contact lens near the limbus of the artificial sclera–cornea membrane, where deformations are strongest (Fig. 4(a))[32]. The grating lines were oriented radially so that the sensor is sensitive to tangential strain. The contact lens with the transferred DFB laser was then mounted onto the eye model. The grating was pumped at 510 nm, and the laser emission, separated from the pump beam by a long-pass filter, was collected by a fiber and analyzed with a spectrometer, while the pressure was varied through the eye model and monitored with an external manometer (Fig. 4(b)). A clear red shift of the laser line is observed as the pressure increases, confirming the fundamental operating principle of our measurement approach (Fig. 4(c)). However, the measured peak width is limited by the 0.74 nm resolution of the spectrometer; the intrinsic laser linewidth is therefore not extracted from these measurements. To partially circumvent this limitation, we fitted the observed peaks with a Lorentzian profile, which allowed us to locate the peak position with a precision much higher than the instrumental linewidth itself. In this way, we tracked the wavelength–pressure dependence over a loading–unloading cycle (Fig. 4(d)) to calibrate our sensor. The calibration curve is linear, with a sensitivity of 0.027 nm/mmHg. This value falls somewhat below the predicted range. This is the direction expected for an estimate that neglects the Poisson compression of the layer thickness and assumes an ideal transfer of the deformation from the membrane to the lens, and the agreement therefore remains reasonable at the order-of-magnitude level at which the estimate was made. The root-mean-square deviation of the true pressure from the linear fit is 1.2 mmHg. This residual characterizes the sensor together with the artificial eye and the pressure system rather than the sensor alone, and it is of the order of the precision required for tonometric monitoring. On the other hand, the source of this error deserves discussion. First, a systematic shift is observed between the points corresponding to the loading and unloading branches. This hysteresis-like behavior contributes a systematic offset of approximately 1.7 mmHg. We associate it primarily with the artificial eye and the pressure system: we suppose that part of it arises from the thermal drift of the pressure baseline described in Section 2.6, and viscoelastic relaxation of the PDMS membrane may contribute as well. Our data do not allow these contributions to be separated, and a contribution from viscoelastic relaxation of the lens and of the dye film cannot be excluded either. Separating these contributions, and establishing the intrinsic noise floor and the device-to-device variation of the sensor, will require fixed-pressure repeatability measurements, repeated loading cycles on several devices and dwell tests; until then, the residual reported here characterizes a demonstration rather than a sensor specification. Additional bottlenecks for precision are the finite resolutions of the spectrometer and the manometer. While the manometer is an auxiliary element required only for device testing, the spectrometer is an inherent part of the sensor. Clearly, to improve precision without a significant increase in cost or setup complexity, specialized solutions - discussed below - are required. Overall, our DFB sensor already achieves a small calibration residual on the artificial eye and, more importantly, has a much higher potential that is currently limited mainly by auxiliary equipment. It is useful to express this residual as a strain precision, for comparison with the bound discussed above. With $d\lambda/d\varepsilon \approx 346$ nm, a residual of 1.2 mmHg corresponds to $\sigma_\lambda \approx 0.032$ nm and hence to $\sigma_\varepsilon \approx 9\times10^{-5}$, while the uncertainty-principle bound for a passive grating of the same nominal aperture, a/L with L = 5 mm, is $6.7\times10^{-5}$. The precision demonstrated here is thus of the same order as that bound and does not exceed it, and the advantage of the present scheme should accordingly be regarded as a potential enabled by the narrow lasing line rather than as an achieved result. We note that a/L evaluated with the full aperture is the most favorable value a passive grating of this size could attain: the length over which the period remains correlated is reduced by the defects visible in Fig. 3(a) and by finer disorder, and a coherent length below about 3.7 mm would already place the present result beyond the bound applicable to our own structure. This length is difficult to determine, but the macroscopic defects visible even by eye in the photograph in the inset of Fig. 4(a) indicate that it is shorter than the full aperture; as it has not been measured, we do not base a claim on it.

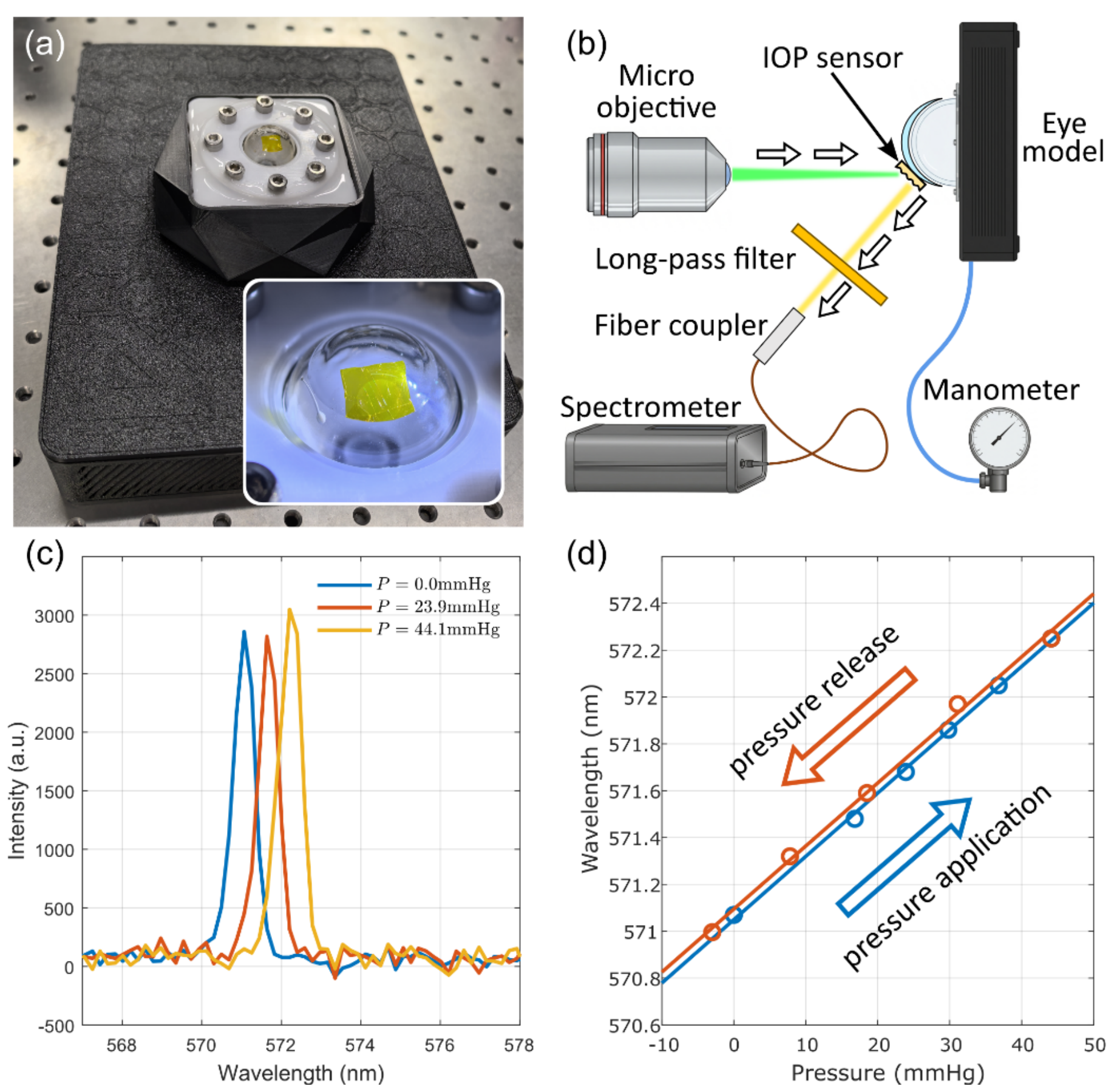

Fig. 4 (a) Photograph of the artificial eye model with a contact lens and DFB sensor placed on top; the inset shows the sensor at higher magnification. (b) Schematic of the measurement: the DFB structure is pumped through an external micro-objective at 510 nm wavelength; the laser emission, after long-pass filtering, is collected by a fiber coupler and analyzed with a spectrometer. Simultaneously, the eye model applies pressure to the artificial membrane, monitored by an external manometer. (c) DFB laser emission spectra demonstrating narrow, spectrometer-limited peak width. (d) Calibration curve of the DFB-laser-based IOP sensor showing the emission wavelength as a function of applied pressure. The calibration residual is primarily limited by the pressure measurement accuracy and hysteresis effects. Negative gauge values on the release branch are attributed to a thermal drift of the pressure baseline (see Methods), not to the optical readout.

## 4. Discussion

Several issues deserve separate discussion. The first is eye safety. Because the pump light is delivered to the eye, the appropriate framework is a per-aperture comparison against the ANSI Z136.1 / IEC 60825-1 laser-safety standards. A reading requires a single 8 ns pump pulse at 510 nm (each calibration point in Section 2.6 is a single-pulse spectrum), producing a corneal radiant exposure of $H_{cornea} \approx 1$ mJ/cm² over a ≈5 mm spot. A diffuser placed before the eye gives an approximately top-hat intensity profile, so the single-pulse fluence represents both the average and, to within the residual non-uniformity, the peak value. Both the pump (510 nm) and the DFB emission (571 nm) lie in the retinal-hazard region (400–700 nm), where the nanosecond exposure limit carries no wavelength correction; the closest available in-vivo threshold data (532 nm, 7 ns) therefore apply directly[51–53]. Crucially, the pump is focused at $z \approx 17$ mm in front of the cornea and enters the eye as a diverging cone. The virtual source thus lies well inside the near point of accommodation (~100 mm) and cannot be imaged onto the retina. From the paraxial vergence relation, the object vergence is $V = -1/z \approx -59$ D, whereas a maximally accommodated eye adds at most ~+10 D. Propagating the pump beam through the eye with the paraxial ray-transfer (ABCD) formalism, with the eye modeled as a thin lens ($f \approx 14.5$ mm at maximal accommodation) followed by the air-equivalent length $d = 17$ mm, and neglecting both the diffuser, which only reduces the beam brightness, and clipping by the pupil, gives a minimal retinal spot diameter $d_{retina} = 2(w_0/z_R)[(Az_R)^2 + B^2]^{1/2} \approx 3.7$ mm, where $w_0 \approx 1.3$ µm and $z_R \approx 10$ µm are the waist radius and the Rayleigh range of the pump beam, $A = 1 - d/f$ and $B = z + d - zd/f$; the same model gives a beam diameter $d_{cornea} \approx 4.4$ mm at the cornea, slightly below the measured ≈5 mm. Assuming full transmission to the retina, the retinal radiant exposure is therefore at most $H_{retina} = H_{cornea}(d_{cornea}/d_{retina})^2 \approx 1.4$ mJ/cm². This retinal exposure is about an order of magnitude below the measured retinal-injury threshold of ≈13 mJ/cm², obtained from the large-spot branch of the in-vivo data (where the threshold radiant exposure is constant and our ≈4 mm patch lies), with the caveat that this is a margin to the median damage threshold rather than a normative MPE margin. Because the readings are single-pulse and infrequent (e.g. hourly), successive measurements are separated by intervals far exceeding any thermal-accumulation window, so the repetitive-pulse reduction ($C_p = N^{-1/4}$) does not apply and each pulse is treated independently. Referenced instead to the small-source MPE — which credits the eye with a focusing capability it does not have for this diverging source — the energy would formally appear to exceed the limit; this is an artifact of the small-spot branch, whereas our source is physically confined to the large-spot regime. The DFB emission at 571 nm is far weaker still. It is at the nanojoule level (~$10^{-3}$ µJ) and, owing to the limbal placement of the sensor, is not directed into the pupil, so the retinal-hazard analysis does not apply to it; the relevant surface exposure is ≈$5\times10^{-9}$ J/cm², roughly two orders of magnitude below the corneal-plane limit even under the conservative assumption of ocular incidence. Because the grating is one-dimensional, this emission is also strongly astigmatic and does not form a collimated beam, which only reduces any hypothetical retinal irradiance further. In practical terms, a single-pulse reading at the present fluence is therefore already safe with a comfortable margin: a working device would keep the single-pulse corneal fluence at or below the ≈1 mJ/cm² level used here. This margin can be widened further by optimizing the DFB structure.

Another important issue is the full integration of the active sensor components within the lens. One aspect is pump integration. Despite the relatively high operating fluence of our device (≈1 mJ/cm²), threshold fluences roughly four to five orders of magnitude below our present value are well established for organic DFB lasers. In particular, lasing at ≈0.02 µJ/cm² has been reported using a pump spot of $5.2 \times 10^{-3}$ cm² (≈0.8 mm) [54], corresponding to a pump energy of only ≈0.1 nJ, i.e. a peak power of ≈13 mW over an 8 ns pulse. This demonstrates considerable potential for grating optimization to reduce the lasing threshold and enhance energy efficiency. For instance, increasing the modulation depth would strengthen mode coupling, reducing the hybrid-mode group velocity further and improving optical feedback. The grating design can also be refined: a mixed-order grating supporting second-order diffraction can enhance mode coupling without introducing additional out-of-plane losses, leading to a corresponding increase in the quality factor and a decrease in the group velocity of the BIC mode. Alternatively, a π-shifted grating with a built-in cavity can operate at the anti-crossing point at the boundary of the Brillouin zone and emit directly into a waveguide mode, resolving mode competition and avoiding out-coupling losses, thereby lowering the lasing threshold. Commercial InGaN (GaN-based) green laser diodes are available across 505–525 nm with continuous-wave output powers of tens to over a hundred milliwatts in a chip-scale package. Under short-pulse (gain-switched) operation at low duty cycle, their peak power rises well above the CW rating, into the hundreds-of-mW-to-watt range. The pump requirement identified above (~$10^{-2}$ W peak over 8 ns) is therefore met with a substantial margin by existing single-emitter green laser diodes. Because the diode emitter footprint is comparable to or smaller than this pump area, the required fluence can be reached with the source in direct proximity to the DFB film, without imaging optics. Compact pumping of the DFB is thus compatible with present diode technology. The pump itself is not a significant heat source: with one single-pulse reading per hour, even the present pump energy of ≈200 µJ corresponds to an average power of about $6\times10^{-8}$ W, and even if the pulse were fully absorbed in the dye film, the heat would leave the film within about a microsecond and, spread over the adjacent 100 µm of the lens, would raise its temperature by only about 0.1 K.

Another aspect is the integration of a miniaturized spectrometer, especially given our extremely high demands on its resolution. The relevant figure of merit here is the bandwidth-to-resolution product, $N = BW/\delta\lambda$, i.e. the number of independently resolvable spectral channels, which directly reflects the technological complexity of a spectrometer. Every dispersive, interferometric, and resonant architecture is subject to a well-known resolution–bandwidth trade-off, in which the achievable N is bounded by the cavity finesse or the corresponding free-spectral-range constraint [55–57]. State-of-the-art benchtop optical spectrum analyzers operate at $N \approx 10^5$ (e.g. $\delta\lambda$ = 0.01 nm over a 1500 nm window). In our case, a calibration-free readout must then cover BW ≈ 2.2 nm (the ≈0.5 nm pressure-tuning window plus the device spread), corresponding to $N \approx 220$ resolvable channels at $\delta\lambda$ = 0.01 nm — still more than two orders of magnitude below the resolution–bandwidth limit of current technology. If instead a one-time per-device wavelength calibration is performed, the device term is removed, and the window reduces to the ≈0.5 nm tuning range ($N \approx 50$). Achieving this resolution within such a narrow window is routinely accomplished without recourse to long optical paths, high-finesse cavities, or computational reconstruction. Several approaches are therefore available. The most straightforward employs a commercially available bandpass filter with a sharp spectral edge or a high-Q resonator; a small shift of the laser line produces a large change in transmitted intensity, which serves as the readout signal. Another promising route is based on random spectrometers, which exploit the sharp spectral features of random interference in disordered structures. These devices force light to undergo multiple random round trips, accumulating large phase differences within a small volume and achieving remarkably high resolution in ultra-compact dimensions.

A further consideration is that the lasing wavelength responds to any perturbation of the grating period or effective index, so cross-sensitivities must be considered. The dominant thermal contribution is the thermal expansion of PDMS: with a thermal expansion coefficient of $1.8\times10^{-4}$ K$^{-1}$ [58] and the same conversion factor as for pressure, it corresponds to about 0.06 nm/K, or roughly 2 mmHg per kelvin on the present calibration. The F8BT layer, by contrast, remains glassy over this range ($T_g$ = 130 °C [59]) and expands considerably less, which may attenuate the effect; the strain of the PDMS, however, reaches the grating in the same way as the pressure-induced one and cannot be neglected. The difference between room and body temperature is a constant offset that is removed by calibration, but temperature variations during wear are not, and a practical device must therefore measure temperature together with pressure. One route fits the present platform naturally: pressure produces markedly different radial and tangential strains at the sensor position, whereas thermal expansion is close to a uniform scaling and gives nearly equal components, so two gratings oriented along these directions would allow both quantities to be recovered from the pair of wavelength shifts. Hydration acts similarly but enters as a constant offset: PDMS swells only weakly in aqueous media and reaches equilibrium shortly after contact with the tear film, so the effect is removed by calibrating the hydrated lens.

Operational lifetime is set by photodegradation of the gain layer. For F8BT lasers the dominant mechanisms are chain scission and cross-linking rather than photo-oxidation, the laser shutting down before photo-oxidation becomes apparent[60]. The relevant figure of merit is the accumulated pump dose rather than elapsed time: for an unencapsulated DFB laser based on a comparable conjugated polymer, the output falls to 1/e of its initial value after a dose of ≈6 J/cm² [61]. At the present pump fluence of ≈1 mJ/cm² per reading this corresponds to several thousand readings, against about 720 over the service life of a monthly disposable lens at one reading per hour, and the margin grows as the required pump fluence is reduced. A direct photostability measurement on our own films nevertheless remains necessary.

Finally, biocompatibility and encapsulation must be addressed prospectively. In the present proof-of-concept the F8BT film is exposed at the outer surface, so that leaching, adhesion under blinking and dye toxicity would all need to be controlled before in vivo use. Both constituent materials are favorable in this respect: PDMS is a well-established biocompatible elastomer [62], and F8BT-based conjugated-polymer nanoparticles have shown low cytotoxicity [63]. More importantly, the intended architecture embeds the F8BT layer within the PDMS body rather than leaving it at the surface: full encapsulation mechanically fixes the film against abrasion and delamination during blinking and forms a diffusion barrier that suppresses any leaching pathway. A dedicated study of encapsulation integrity, long-term photostability, and biocompatibility under physiological conditions is planned as the next step. More generally, the temperature dependence, long-term mechanical stability, optical output power and integrated readout of the sensor have not been demonstrated in the present work.

## 5. Conclusion

In this work, we have demonstrated a contact-lens-mounted IOP sensor based on a distributed-feedback laser formed by a surface-modulated ultrathin F8BT dye layer. The key conceptual advance is that, unlike conventional optical strain sensors whose precision is bounded by the spatial uncertainty principle linking grating period and grating size, our device transfers the measurement to the spectral domain: the lasing wavelength is determined by the grating period, but its linewidth is set by the temporal coherence of the lasing mode and is therefore disproportionately narrow relative to the millimeter-scale footprint imposed by the contact lens. By engineering the surface modulation of the slab waveguide to operate at a symmetry-protected bound state in the continuum at the Γ point, we obtain a high-Q dark mode that wins mode competition, lases stably, and emits close to the grating normal, simplifying readout.

Equally important, the entire DFB structure is fabricated by a simple, scalable, and polymer-compatible route - UV holographic lithography of a positive photoresist followed by replication into the F8BT waveguide and float-off transfer onto PDMS. This process forms the patterned active layer in a single spin-coating step on the master and transfers it onto the lens by float-off, avoiding the separate imprint-and-transfer stages otherwise needed to pattern sub-wavelength features in the active layer on a soft, curved final substrate.

Tested on an artificial eye whose pressure-induced deformation is comparable in scale to that reported for the human eye, the sensor delivers 0.027 nm/mmHg sensitivity and a 1.2 mmHg calibration residual under phantom conditions, already within the range relevant to tonometric monitoring. The remaining error budget is, in our assessment, dominated by auxiliary equipment (spectrometer and manometer resolution, and the hysteresis between the loading and unloading branches, which we associate primarily with the artificial eye and the pressure system), rather than by the sensing principle itself. Combined with clear paths toward grating optimization (mixed-order or π-shifted designs), miniaturized wavelength-meter readout (edge filters, high-Q resonators, random spectrometers), and full lens integration, with the passive DFB sensor read out by an on-lens wavelength meter and pumped either by an integrated compact laser diode or by an external reader at the moment of measurement, our approach opens a realistic route to wearable, continuous, and minimally invasive intraocular-pressure monitoring.

**Acknowledgments.** We acknowledge Prof. S. Dyakov and Prof. N. Gippius for providing the code for the Fourier Modal Method used in the optical calculations.

**Disclosures.** The authors declare no competing interests except for the patent application PCT/IB2025/058053 filed 07/08/2025 associated with this work and held by XPANCEO RESEARCH ON NATURAL SCIENCE L.L.C.

**Data availability.** The datasets generated and analyzed during the current study are available from the corresponding author upon reasonable request.

**Supplemental document.** See Supplementary materials for supporting content.

# Supplementary materials
# "Contact lens with stretchable distributed-feedback laser for intraocular pressure monitoring"

SERGEI A. IVANOV,[1] ILIA M. FRADKIN,[1,*] EKATERINA S. MUSIKHINA,[1] ROMAN V. KIRTAEV,[1] ALEKSANDR A. KHREBTOV,[1] ANDREY A. VYSHNEVYY,[1] ALEXANDER A. MARCHENKO,[1] VALENTIN R. SOLOVEI,[1] ILYA P. RADKO,[1] ALEKSEY V. ARSENIN,[1] AND VALENTYN S. VOLKOV[1]

[1]*Emerging Technologies Research Center, XPANCEO, Dubai Investment Park First, Dubai, UAE*
**fradkinim@xpanceo.com*

### Supplementary Note 1: Mechanical deformations of artificial eye

The sensitivity of the artificial eye model to applied pressure is determined primarily by its thickness and the stiffness of the polymer material. The known mechanical properties of PDMS are close to those reported for the tissues of a real eye, allowing us to fabricate the sclera–cornea membrane with thicknesses similar to those of the corresponding anatomical layers. However, the Young's modulus of a particular PDMS sample can vary significantly depending on the components, their ratio, and the polymerization conditions. We therefore first tested membranes of different thicknesses to obtain the required stiffness. The actual Young's modulus was then extracted by comparing experimental deformations with numerical simulations. Here we present the measurements and simulations for this specific structure. It is instructive to compare the membrane stiffness $E{\cdot}t$, which governs the tangential strain under pressure and therefore removes the ambiguity associated with the individual values of the Young's modulus and the thickness. For the corneal region of our membrane, $E{\cdot}t \approx 1.47$ MPa × 450 µm ≈ 0.66 kN/m, while in vivo measurements of the human cornea give a tangent modulus of 0.54-0.71 MPa at a central thickness of about 540 µm [1], that is $E{\cdot}t \approx$ 0.29-0.38 kN/m. Our artificial cornea is therefore of the same order of stiffness as a real one. It should be kept in mind, however, that the mechanical response of an eye cannot be reduced to a single number, and that the outcome of such a comparison depends on which characteristic is chosen. The tangential strain in the limbal zone, the change of curvature at the corneal apex, the indentation stiffness of the central cornea and the membrane stiffness $E{\cdot}t$ all describe the same object, yet a given sample may appear stiffer than a reference by one of them and softer by another. Part of the spread between the comparisons made below therefore reflects the choice of the characteristic rather than a genuine difference in stiffness, and we use these comparisons only to establish that our model reproduces the correct order of magnitude of the deformations.

First, we applied pressure to the membrane alone (without the lens) and acquired profile photographs to estimate the deformations. As shown in Fig. S1(a), we superimposed translucent circles corresponding to the corneal and scleral curvatures onto the photographs. Importantly, the sizes and relative positions of these circles were taken from the original design; they were calibrated beforehand using scale marks photographed in the same frames. The fabricated structure closely reproduces the designed shape, and pressure-induced deformations are clearly detectable. Naturally, the tiny local changes in curvature at specific points are too small to be reliably parametrized from photographs. Nevertheless, the elevation of the corneal apex is relatively large because it accumulates the deformations of the entire membrane. In particular, at a pressure of 40 mmHg the apex displacement reaches approximately 150 µm (Fig. S1(a)). Elevations at lower pressures were also measured with a precision of about 10 µm (Fig. S1(b)). By comparing these data with simulation results, we conclude that the Young's modulus of PDMS resulting from our fabrication process is $E \approx 1.47$ MPa. As Fig. S1(b) shows, this value yields good agreement between the experimental measurements and the finite-element calculations performed in COMSOL Multiphysics.

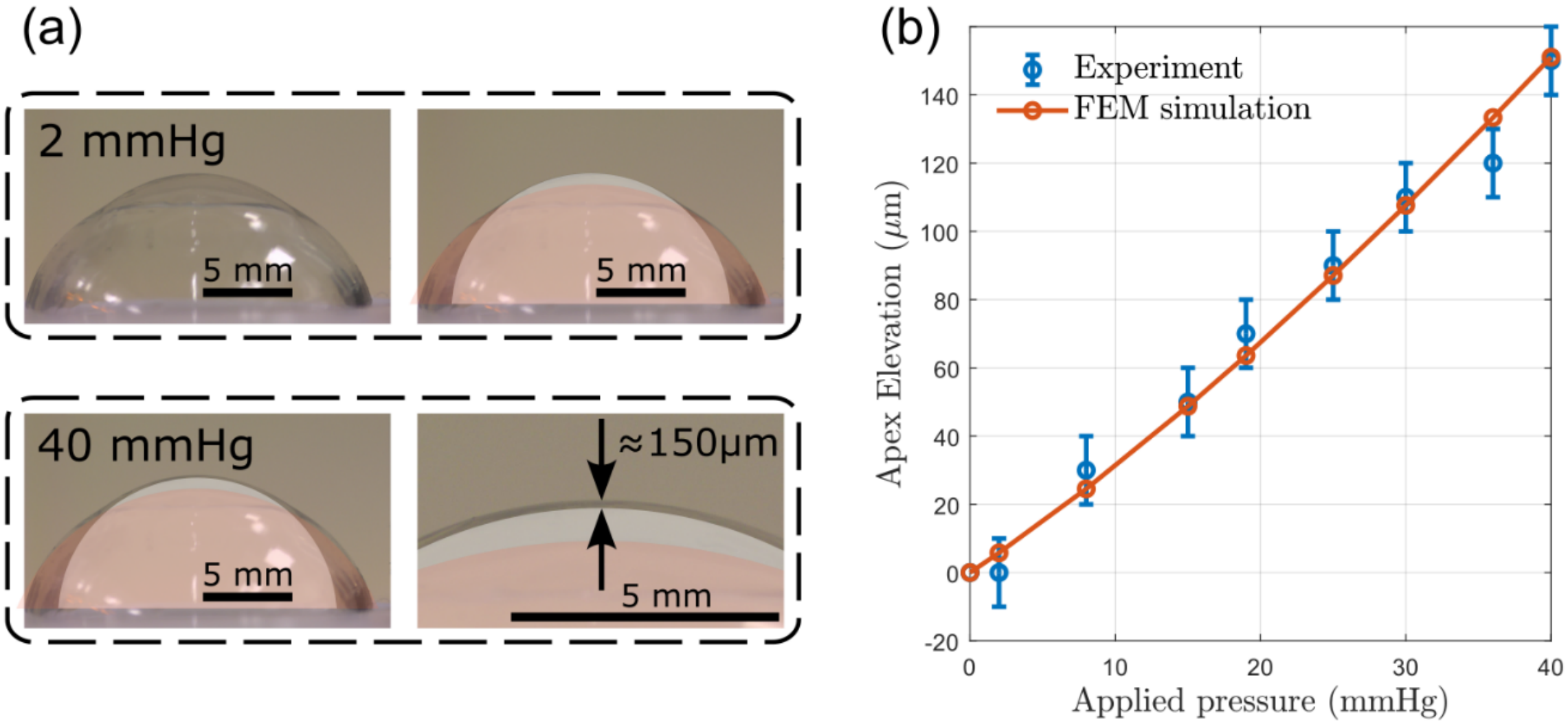

Fig. S1 (a) Profile photographs of the artificial sclera–cornea membrane under different pressures. Translucent red and white circles, which almost perfectly match the membrane shape, correspond to the nominal radii of the sclera and cornea used in the design. The corneal apex is elevated by approximately 150 µm under a pressure of 40 mmHg. This elevation allows the stiffness and Young's modulus of the membrane to be estimated. (b) Comparison of experimentally measured and numerically simulated (E = 1.47 MPa) elevations of the corneal apex.

Once the Young's modulus is known, we can simulate all deformation scenarios: the membrane alone, the membrane with a lens sliding on its surface, and the membrane with a lens fully adhered to it. We begin with the bare membrane. Figure S2(a) shows the volumetric strain at 40 mmHg as a deformation indicator. The largest strains are concentrated in the limbal area and exceed those in the adjacent corneal and scleral regions. This clearly demonstrates that, at least in our artificial model, applied pressure does not simply scale the entire structure uniformly but also changes its shape. Although the behaviour of our model may differ quantitatively from that of a real eye, similar effects may be relevant there as well. In this context, the corneal radius commonly reported in the literature is not a uniquely defined quantity: the local radius of curvature at a given point on the cornea and the mean corneal radius are generally different.

We therefore characterize the deformation of our membrane using two quantities: the relative change in the radius of curvature at the corneal apex (center) and the transverse strain in the limbal zone. We specifically choose the transverse component because it is nearly homogeneous across the membrane cross-section, in contrast to the other components, and because it is predictably transferred to the lens. This transverse strain determines the change in the corresponding circumference, which is the key parameter for shape change. The limbal zone is of the greatest interest because the transverse deformations relevant to our sensor are highest there.

We first examine the deformation of the bare membrane without any lens (Fig. S2(b)). The deformations are linear over the entire pressure range. The relative change in curvature at the corneal apex is much larger than the limbal strain. However, a large local change in curvature near the apex does not imply correspondingly large material deformations, as it mainly reflects a high curvature gradient rather than large absolute strains. The limbal strain itself is somewhat larger than the typical eye deformations reported in the literature, which are qualitatively indicated by the light-green band. By this particular measure, our artificial membrane is slightly softer than a typical eye, whereas by its membrane stiffness E·t it is somewhat stiffer, which illustrates the ambiguity discussed above.

Ultimately, we are interested in measuring deformations via the lens placed on the membrane. We consider two limiting cases: a freely sliding lens (in the presence of a lubricant, Fig. S2(c)) and a no-slip (fully adhered) lens (Fig. S2(d)). In both cases the lens is assumed to be in full contact with the membrane. As seen in Fig. S2(c), even a sliding lens adds some rigidity to the system. The limbal strain decreases only slightly and may be neglected, but the change in curvature at the corneal apex is suppressed by a factor of two. The reason is that the curvature change at the apex is only weakly linked to the actual material deformations; it can change rapidly even for very small strains and therefore cannot serve as a reliable benchmark. At the same time, the transverse strain in the lens at a point opposite the limbus is more than three times smaller than the limbal strain inside the membrane. This effect is clearly due to the ability of the lens to avoid excessive stretching by sliding along the membrane surface.

We therefore finally compute the same deformations for the adhered lens (Fig. S2(d)), which corresponds to the practical case in the absence of lubricant. Here, stress can no longer be relieved by sliding, and the lens truly stiffens the whole system by taking up part of the load. Indeed, the strains in the membrane and in the lens become much closer, and both fall within the green target area. From this we conclude that the adhered system of artificial sclera–cornea membrane and contact lens approximately reproduces the stiffness of the real eye. Admittedly, there are significant approximations, as discussed above, and the model reproduces the scale of the pressure-induced deformations rather than the eye itself. Given the large uncertainty regarding the true deformations of the human eye under IOP changes, we consider it adequate for the purpose of testing the sensor, while noting that it cannot substitute for measurements under more realistic conditions.

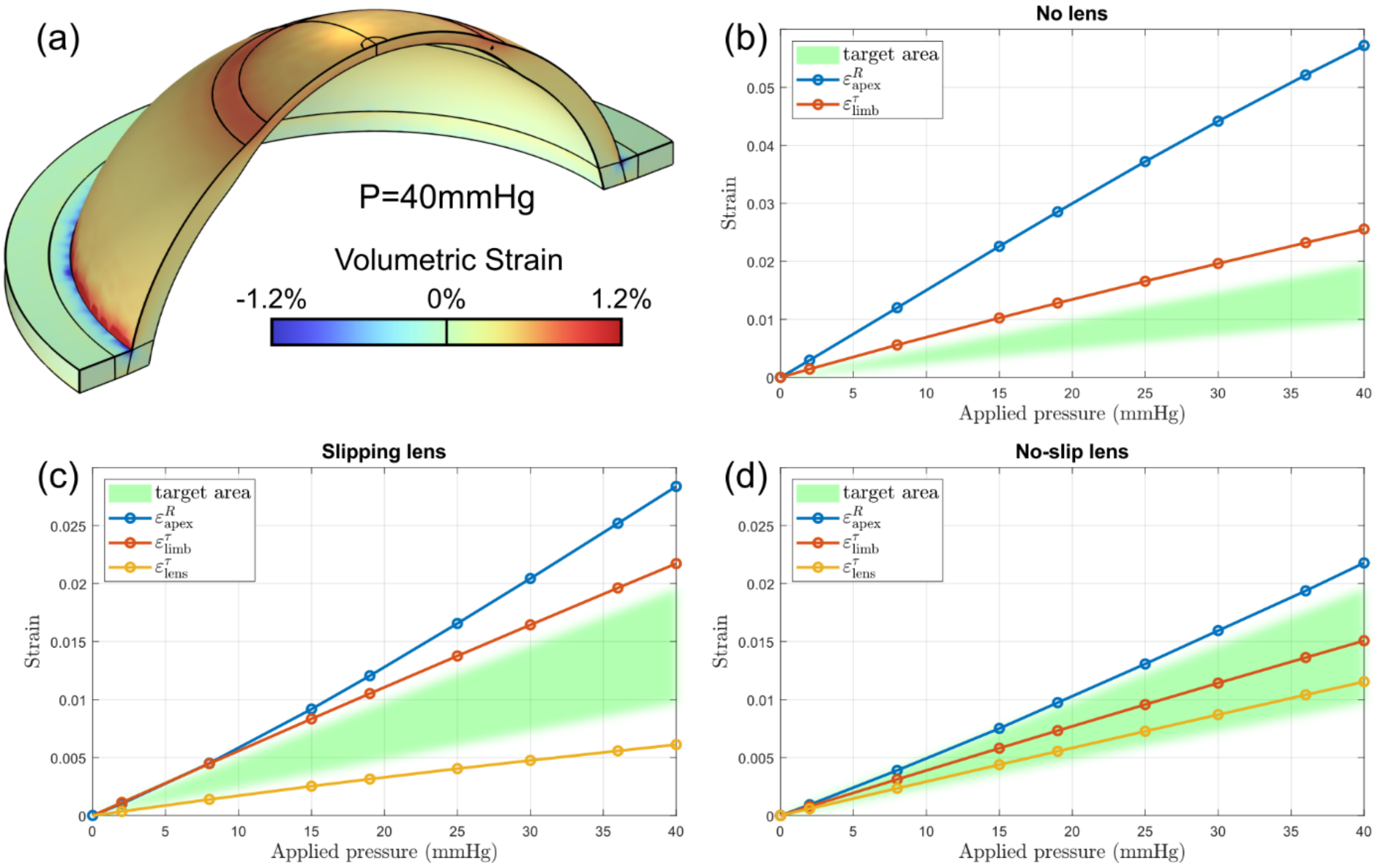

Fig. S2 (a) Volumetric strain distribution in the artificial sclera–cornea membrane. Strain in the limbal area between the sclera and cornea exceeds that in the surrounding regions. (b–d) Dependence of strain on applied pressure for (b) the bare membrane, (c) the membrane with a sliding contact lens on top, and (d) a no-slip lens adhered to the membrane. Blue lines show the relative change in the radius of curvature at the corneal centre, red lines correspond to the transverse strain in the limbal zone of the membrane, and yellow lines correspond to the transverse strain inside the lens at a point opposite the limbus. The light-green band indicates the typical corneal strains reported in the literature.

## Supplementary Note 2: Dye emission spectrum

The emission spectrum of the dye layer is a crucial property of our structure. It determines the required wavelength of the resonator and, consequently, all the geometrical dimensions of the structure. As seen in Fig. S3, the intensity of spontaneous emission is maximal in the range of approximately 530–590 nm. We therefore designed our resonator to operate near 570 nm, which lies within this range and corresponds to the absence of dye absorption.

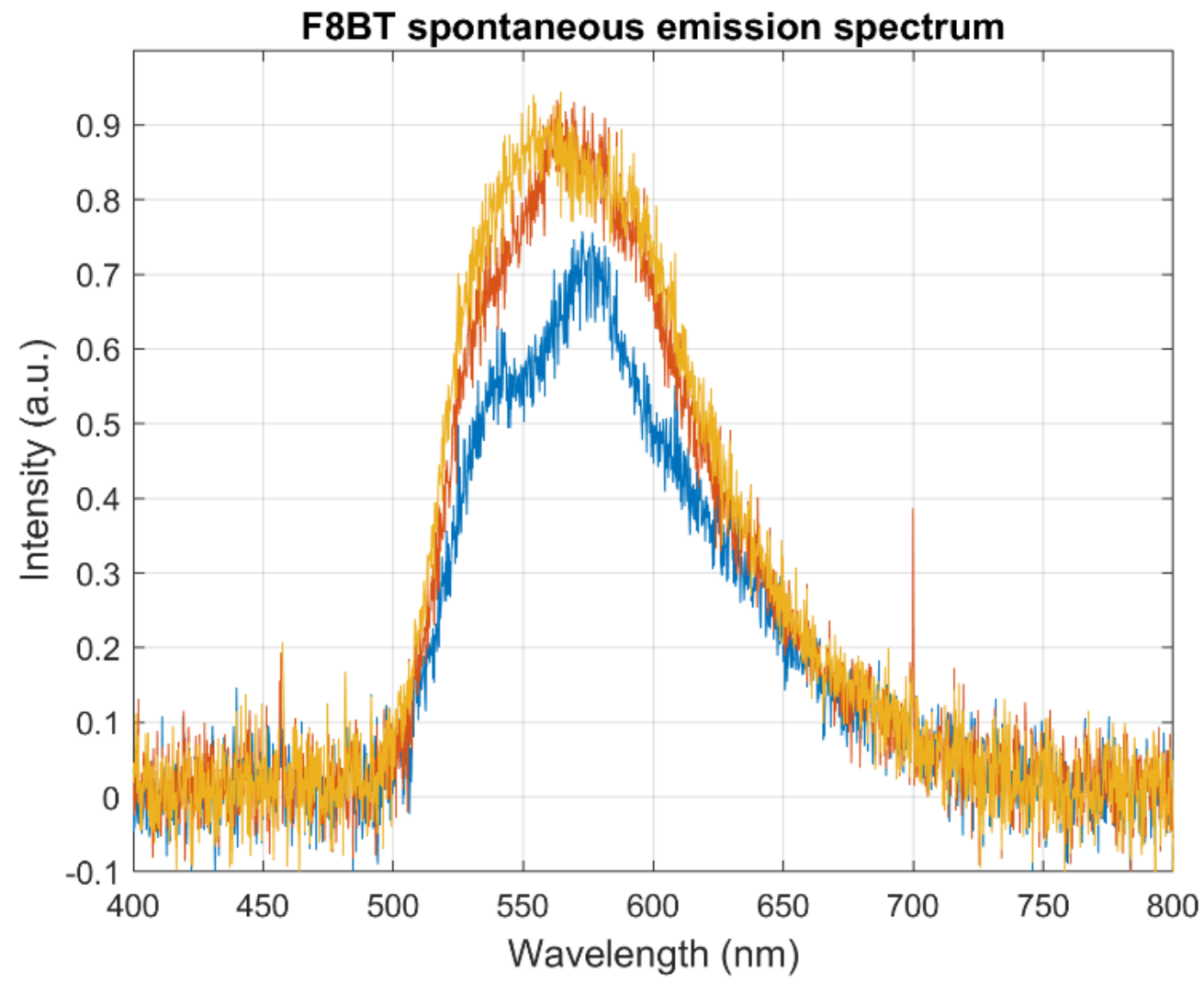

Fig. S3 Fluorescence spectra of the F8BT dye measured from different samples. The emission is maximal in the range of ~530–590 nm.

## Supplementary Note 3: Dye spin curve

This section presents the spin curve demonstrating the dependence of the final F8BT dye film thickness on the spin-coating speed (Fig. S4). The films were deposited from a 25 mg/ml F8BT solution in toluene. The presented graph serves as a calibration curve, enabling the selection of the optimal spin speed to achieve F8BT layer of a precisely specified thickness.

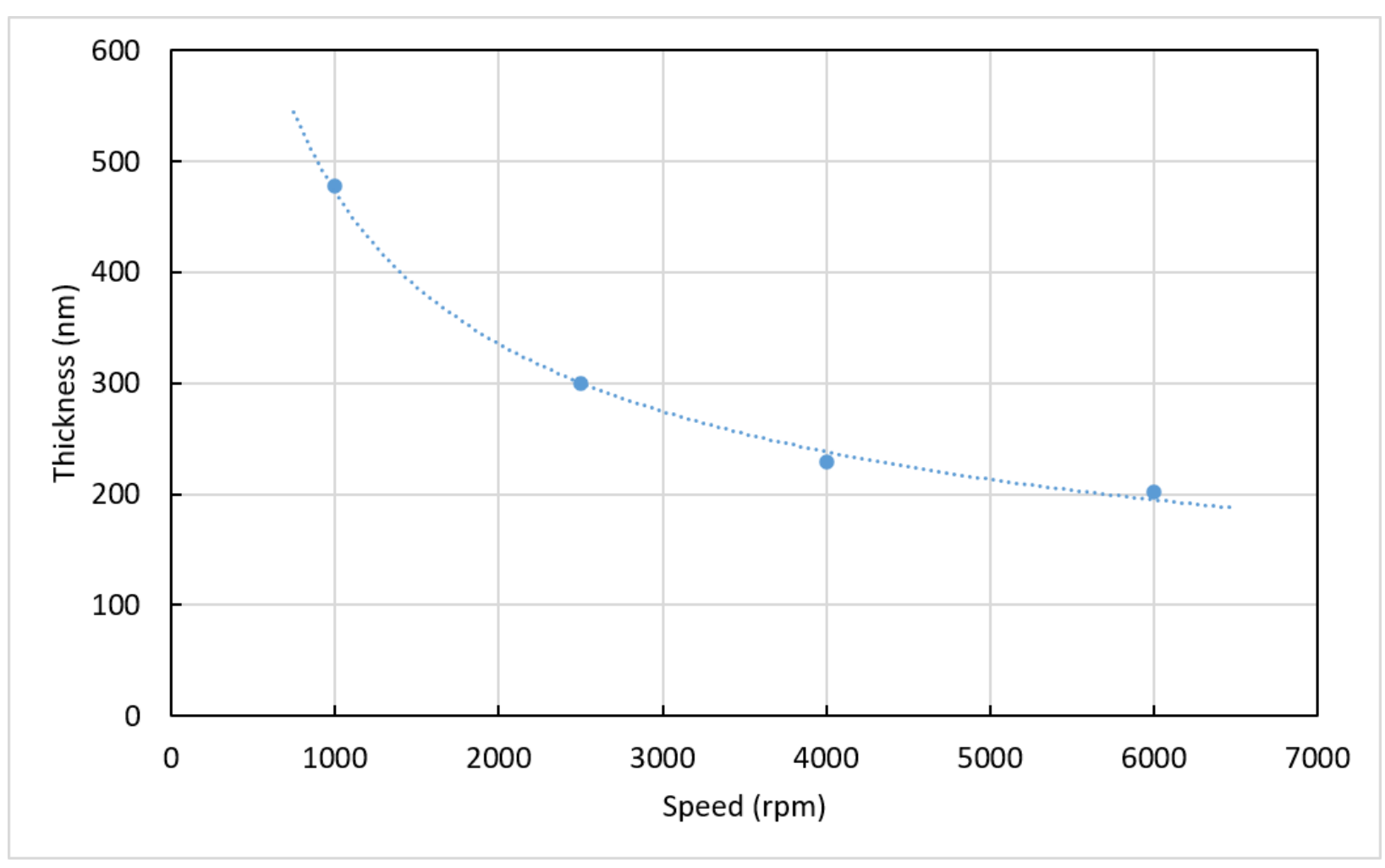

Fig. S4 Calibration curve for thickness of the spin-coated F8BT dye as a function of rotation speed.

## Supplementary Note 4: Dependence of the DFB output on the pump energy

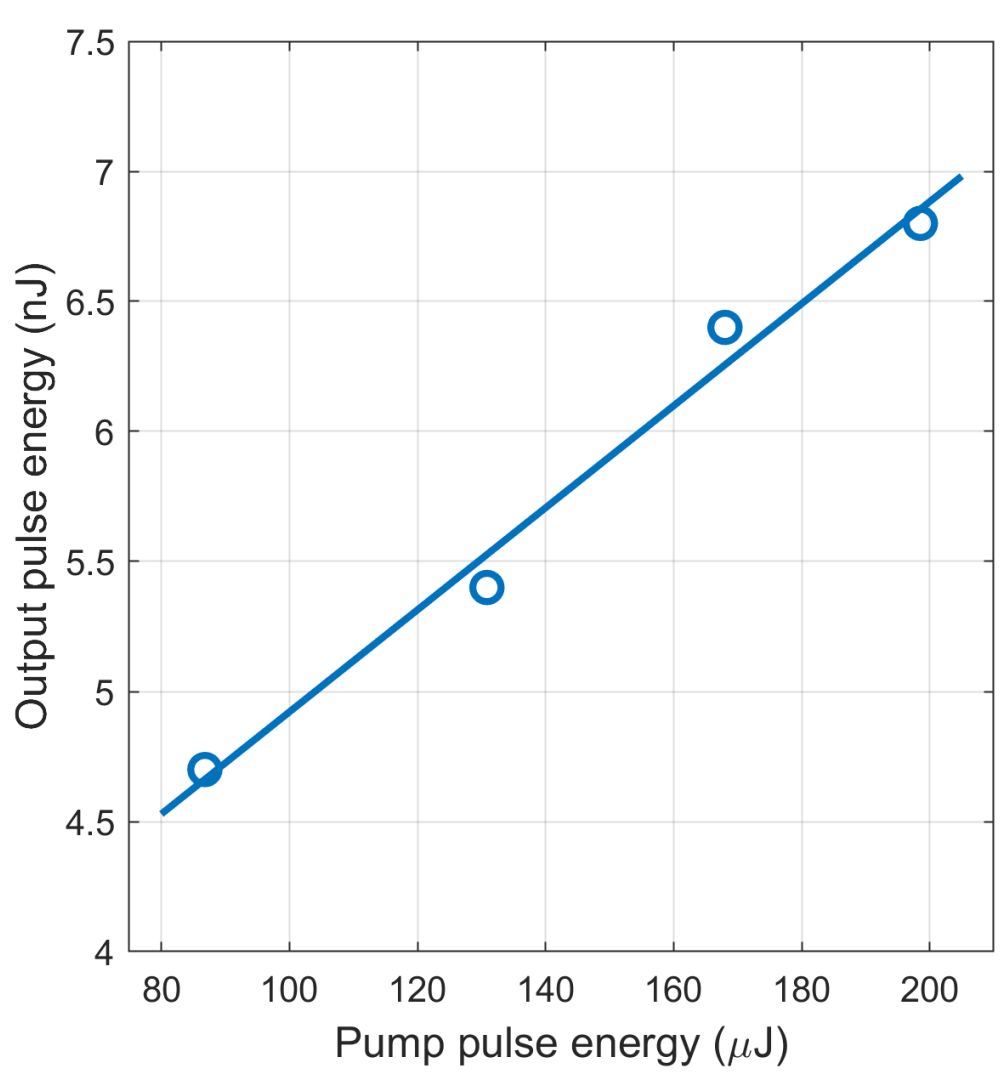

Fig. S5. Output pulse energy of the DFB laser as a function of the pump pulse energy. The line is a linear fit to the measured points, shown as a guide to the eye. Lasing is observed at all pump energies measured, down to 87 µJ, so the lasing threshold lies below this value.

## Supplementary Note 5: Comparison with other contact-lens intraocular-pressure sensors

Table S1 collects the contact-lens sensors against which the present device can reasonably be set. Quantities are given as reported in the original work and in the units used there. We have deliberately not converted them to a common scale: the measured observable differs between approaches - a resistance, a resonance frequency, an angle, a color or a wavelength - and a conversion would require assumptions that the original papers do not support. For the same reason the last column states the principal limitation of each approach rather than ranking them.

| # | Approach | Principle and readout | Reported response | Reported resolution or accuracy | Validation | Principal limitation |
|---|---|---|---|---|---|---|

| 1 | Sensimed Triggerfish [2,3] | Resistive strain gauge; telemetric readout | Output in millivolt equivalents, relative to the first reading of a session | Not given in mmHg; the device does not report absolute pressure | Human; FDA De Novo cleared | Provides a relative signal rather than absolute pressure |
|---|---|---|---|---|---|---|
| 2 | Kim et al. [4] | Resistive strain gauge with an antenna; radio-frequency readout | 0.05 % per mmHg | Minimum detectable change of 0.014 mmHg in bench tests; intraclass correlation of 0.888 against a rebound tonometer | Rabbits; ten human participants | Conductive elements and an antenna on the eye |
| 3 | Zhang et al. [5] | Capacitive sensor; wireless readout | 0.27 MHz per mmHg, equivalently 1121 ppm per mmHg, in a human eye | Not converted to mmHg; linear fit with $R^2$ = 0.91 | In vivo in rabbit, dog and human eyes | Conductive elements on the eye; capacitive readout is sensitive to the dielectric environment |
| 4 | Xiao et al. [6] | Parity-time symmetric antenna pair; wireless readout | 47.31 Ω per mmHg | Not converted to mmHg; $R^2$ = 0.97 against a tonometer in rabbits | Porcine eye in vitro; rabbit eyes in vivo | Conductive elements on the eye |
| 5 | Wei et al. [7] | Contact-lens sensor system; external instrument | Not reported | Mean differences from applanation tonometry within ±2 mmHg; over 80 % of Bland-Altman points within ±5 mmHg and over 60 % within ±3 mmHg | Eighty human eyes | Requires external instrumentation |
| 6 | Agaoglu et al. [8] | Microfluidic channel with a liquid interface; visual readout | Strain on the sensor of 0.007-0.03 % per mmHg depending on the material | Limit of detection of 0.0038-0.0061 % strain, against the 0.04 % strain that the authors associate with 1 mmHg | Enucleated porcine eyes | Readout by imaging of a liquid interface; pronounced temperature cross-sensitivity |
| 7 | Maeng et al. [9] | Photonic crystal with microhydraulic amplification; spectrometer or smartphone camera | ≈ 0.4 nm per mmHg on a silicone eye model; ≈ 0.23 nm per mmHg on a porcine eye | Resolution 0.33 mmHg; limit of detection 3.2 mmHg with a spectrometer and 5.12 mmHg with a smartphone camera | Silicone model eye; porcine eye ex vivo | Broad structural-color band limits the detection threshold |
| 8 | Ding et al. [10] | Moiré pattern between two gratings; camera | ≈ 0.15° per mmHg | Not converted to mmHg | Silicone model eye | Precision bounded by the number of grating periods |
| 9 | This work | Distributed-feedback laser; spectrometer | 0.027 nm per mmHg | Calibration residual of 1.2 mmHg | Artificial eye model (phantom) | Phantom only; external pump and spectrometer; intrinsic linewidth not resolved |

Table S1. Contact-lens intraocular-pressure sensors. Values are quoted as reported in the cited work and in the units used there.

Two features of the table deserve comment. The first concerns what is delivered to the user. Several of the works listed report a response per unit pressure and a coefficient of determination for the linear fit, but do not convert either into a pressure resolution or into an agreement with a reference tonometer; the commercially available device outputs a signal that is relative to the first reading of a session and does not provide an absolute pressure at all.

The second concerns the origin of the precision, and it bears directly on the argument of this work. It is natural to suppose that a larger response per unit pressure makes a better sensor, but what matters is the response measured against the width of the feature from which it is read. The photonic-crystal sensor listed in the table offers an instructive comparison, since it also encodes pressure in a wavelength. Its reported response, 0.23 nm per mmHg on a porcine eye and 0.4 nm per mmHg on a silicone eye model, is an order of magnitude larger than ours, yet its reported limit of detection, 3.2 mmHg with a spectrometer, is larger than the calibration residual of 1.2 mmHg reported here. The reason is that the structural-color band from which the shift is read is broad, whereas a laser line is narrow. This comparison should be read with care, since a limit of detection and a calibration residual are not the same quantity, but the direction of the effect is clear.

This is the sense in which the present approach differs from the others in the table. Its response per unit pressure is unremarkable and is set by the same Bragg relation as in any grating-based scheme. What the laser changes is the width of the observable, which is governed by the temporal coherence of the mode rather than by the size of the periodic structure, and which is therefore not bounded by the number of periods that can be accommodated within a contact lens.